\documentclass{aastex63}
\pdfoutput=1
\received{December 1, 2020}
\revised{}
\accepted{\today}

\submitjournal{ApJ}

\shorttitle{Distance and mass of the NGC 253 galaxy group}
\shortauthors{Karachentsev, Tully, Anand et al.}

\begin{document}

\title{Distance and mass of the NGC 253 galaxy group}

\correspondingauthor{Igor Karachentsev}
\email{ikar@sao.ru}
 
  \author{Igor D. Karachentsev}
\affiliation{Special Astrophysical Observatory, The Russian Academy of Sciences, 
Nizhnij Arkhyz, Karachai-Cherkessian Republic 369167, Russia}

 \author{R. Brent Tully}
 \affiliation{Institute for Astronomy, University of Hawaii, 2680 Woodlawn Drive, 
Honolulu, HI 96822, USA}
 
 \author{Gagandeep S. Anand}
 \affiliation{Institute for Astronomy, University of Hawaii, 2680 Woodlawn Drive, 
Honolulu, HI 96822, USA}

  \author{Luca Rizzi}
  \affiliation{W. M. Keck Observatory, 65-1120 Mamalahoa Hwy, Kamuela, HI 96743, USA}

\author{Edward J. Shaya}
\affiliation{Astronomy Department, University of Maryland, College Park, MD 20743, USA}

 \begin{abstract}
  Two dwarf galaxies: WOC2017-07 and PGC~704814 located in the vicinity
of the nearby luminous spiral galaxy NGC~253 were observed with the
Advanced Camera for Surveys on the Hubble Space Telescope. Their distances of
3.62$\pm$0.18~Mpc and 3.66$\pm$0.18~Mpc were derived using the tip of the red
giant branch method. These distances are consistent with the dwarf
galaxies being members of the NGC~253 group. Based on the radial velocities
and projected separations of seven assumed dwarf companions, we estimated
the total mass of NGC~253 to be $(8.1\pm2.6) 10^{11} M_{\odot}$, giving a total-mass-to-$K$-luminosity ratio $M_{\rm orb}/L_K = (8.5\pm2.7) M_{\odot}/L_{\odot}$. A notable property of NGC~253 is its declined rotation curve. NGC~253 joins four other luminous spiral galaxies in the Local Volume
with declined rotation curves (NGC~2683, NGC~2903, NGC~3521 and NGC~5055) that together
have the low average total-mass-to-luminosity ratio, $M_{\rm orb}/L_K =
(5.5\pm1.1) M_{\odot}/L_{\odot}$. This value is only $\sim$1/5 of the corresponding ratio for
the Milky Way and M~31.
\end{abstract}


\section{Introduction} \label{sec:intro}

Bright spiral galaxies in Sculptor constellation form a diffuse association 
marked by de Vaucouleurs (1959) and Arp (1985). According to Jerjen et al.
(1998) and Karachentsev et al. (2003), these galaxies: NGC~24, NGC~45, 
NGC~55, NGC~247, NGC~253, NGC~300, NGC~7793 and their dwarf companions are
located in a filamentary structure, which extends along the line of sight
from the Local Group to a distance $D\sim7$~Mpc. The Sculptor filament
itself lies in the Local Sheet (Tully 1988), residing also in the Local
Supercluster plane. The central part of the filament is the group of dwarf
galaxies around the luminous spiral NGC~253. At the apparent magnitude of
        $K_s = 3.80^m$ (Jarrett et al. 2003) and the distance of 3.70~Mpc (Anand et al. 2021), the luminosity of NGC~253 corrected
for Galactic and internal extinction is $L_K/L_{\odot} = 10.98$~dex, that exceeds
the luminosity of the Milky Way or M~31.

  In the vicinity of NGC~253 within a radius of 15$\degr (\sim1$~Mpc) there are a
dozen dwarf galaxies with radial velocities close to the radial velocity of
NGC~253 of $V_{\rm LG} = 276\pm2$ km\,s$^{-1}$ (Koribalski et al. 2004). Accurate distances for most of them have been
measured via the luminosity of the tip of red giant branch (TRGB) (Karachentsev
et al. 2003; Cannon et al. 2003; Sand et al. 2014; Toloba et al. 2016). 
Only two assumed dwarf satellites of NGC~253:
WOC2017-07 (Westmeier et al. 2017) and 2DFGRS-S431Z = PGC~704814 (Colless et al. 2003) have not had
reliable distance estimates. In $\S$ 2 we present estimates of their distances, 
made by the TRGB method from the images of these galaxies obtained with the
Advanced Camera for Surveys (ACS) on the \textit{Hubble Space Telescope (HST)}. The distances measured by us confirm
the association of both galaxies with the NGC~253 group.
In $\S$3, updated information on the group membership is used to evaluate the mass of the NGC~253 group.

\section{TRGB distances to WOC2017-07 and PGC 704814}

\begin{figure}[h]
\plotone{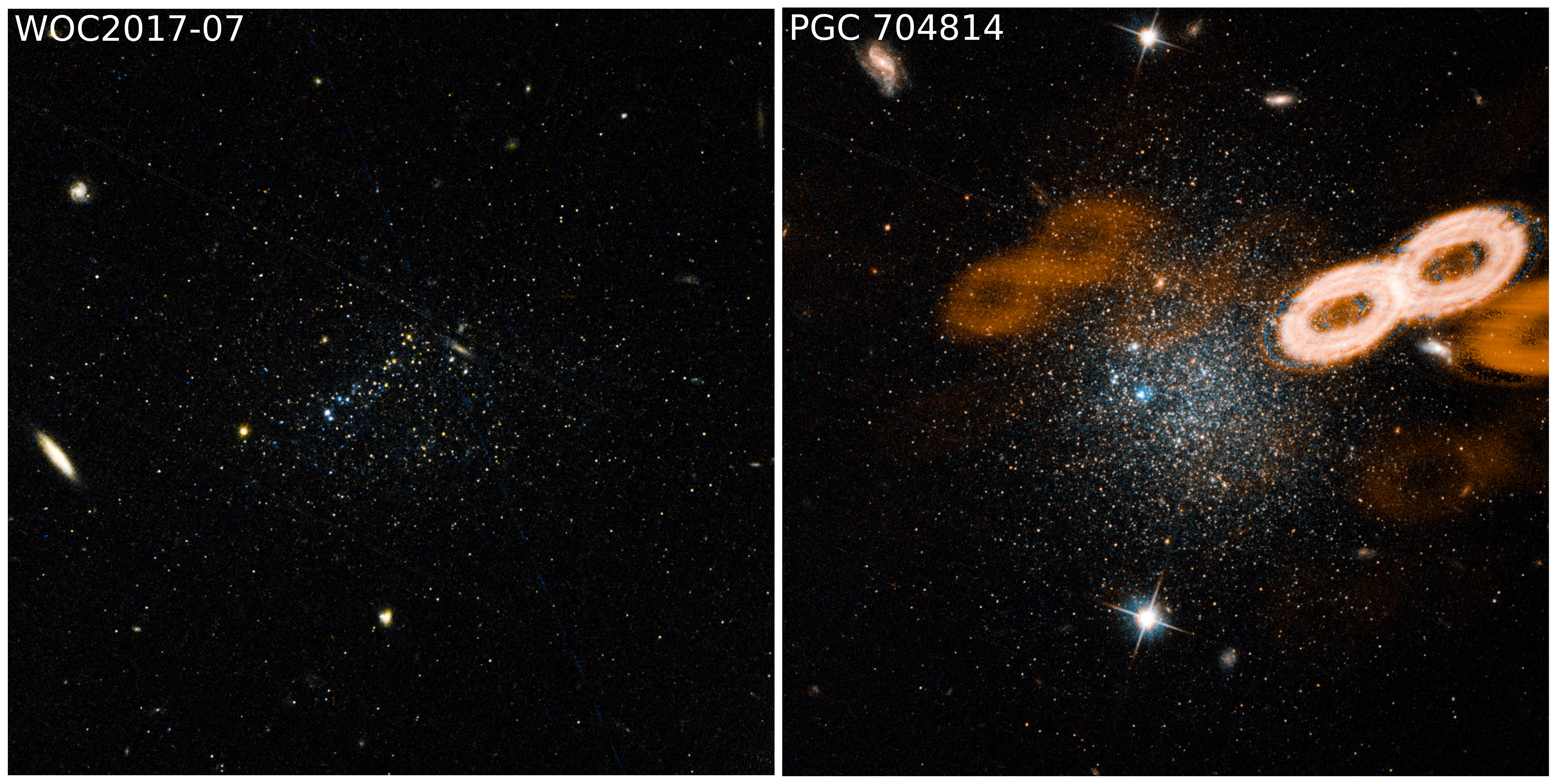}
\caption{HST/ACS combined images of WOC2017-07 and PGC~704814. Each image size is $\sim$ $52\times52$ arcseconds. North is up and east is left. The color images are composites of images with the F606W and F814W filters. The artifacts in the image of PGC~704814 are ghosts from a bright star in the field.}
\label{colorImages}
\end{figure}

We obtained \textit{HST} ACS imaging of WOC2017-07 and PGC~704814 in both F606W and F814W bands (760s each) as part of the Every Known Nearby Galaxy survey (SNAP-15922, PI R. Tully). Color cutouts of these two galaxies produced with this data are shown in Figure \ref{colorImages}. Both galaxies contain visible young and older stellar populations. WOC2017-07 is irregular and lacks clear definition, whereas PGC~704814 is roughly spherical in shape with at least one very prominent young star cluster. The image for PGC~704814 contains numerous artifacts (``figure-eight ghosts"), that are caused by an extremely bright foreground star located in the ACS field of view (but not shown in the cutouts in Figure \ref{colorImages}). For each galaxy, we used DOLPHOT (Dolphin 2000, 2016) to produce PSF photometry with the \textit{.flc} images, using the drizzled F814W image as the alignment reference frame. We cull these photometric catalogs to ensure only resolved sources of the highest quality remain. For this work, we use the quality cuts modified from McQuinn et al. (2017), selecting stars with a signal-to-noise ratio $\ge$ 5 in both bands, type $\le$ 2 (good or faint star), an error flag $=$ 0 (well-recovered stars), ($\mathrm{Crowd_{F606W}+Crowd_{F814W}}) \le 0.8$, and ($\mathrm{Sharp_{F606W}+Sharp_{F814W}})^{2} \le 0.075$. Since the two dwarf galaxies only take up a small portion of the ACS field of view, we isolate the dwarfs to produce a color-magnitude diagram (CMD) that has reduced contamination from background objects. 

\begin{figure}
\plotone{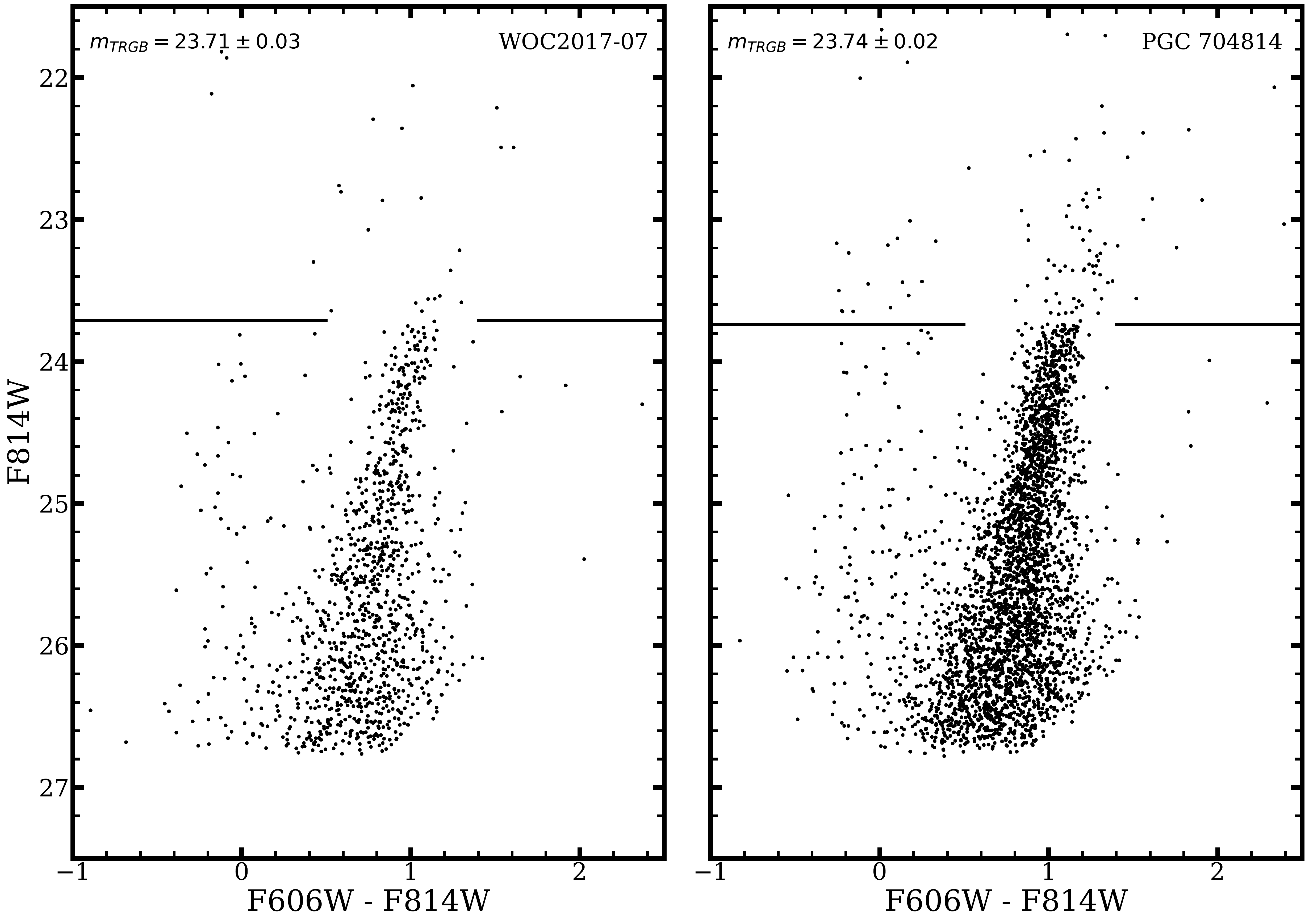}
\caption{Colour-magnitude diagrams of WOC2017-07 and PGC~704814. The TRGB position is indicated by the horizontal line.}
\label{CMDs}
\end{figure}

It is seen in Figure \ref{CMDs} that the main feature present in both of the CMDs is the red giant branch. Both galaxies also have a small population of upper main-sequence stars, indicating there is some ongoing star formation present in both targets. To determine the location of the TRGB, we use the methods described in detail by Makarov et al. (2006) and Wu et al. (2014). Briefly, we perform artificial star experiments with DOLPHOT to quantify the levels of photometric errors, bias, and completeness present in the measured photometry. The results from these experiments allow us to fit the observed luminosity function of red giant branch (RGB) and brighter asymptotic giant branch (AGB) stars with a broken power-law, with the break signifying the location of the TRGB. The physical reason for this parametrization lies in the fact that after undergoing the helium flash and leaving the TRGB, these stars immediately drop down on to the horizontal branch where they are significantly less luminous. This leaves a sharp break on the observed luminosity function at the location of the TRGB-- an in-depth explanation of the method can be found in the original references (Makarov et al. 2006, Wu et al. 2014). We find $m_{TRGB} =$ 23.71 $\pm$ 0.03 for WOC2017-07, and $m_{TRGB} =$ 23.74 $\pm$ 0.02 for PGC~704814. The TRGB in the F814W band is slightly sensitive to the metallicity and age of the underlying stars, so we adopt the absolute magnitude and color calibration of the TRGB presented by Rizzi et al. (2007). Taking into account the (small) foreground extinction (Schlafly \& Finkbeiner 2011), we find $D = 3.62\pm0.18$~Mpc for WOC2017-07, and D $=$ 3.66$\pm$0.18~Mpc for PGC~704814. The reduced photometry, full-field CMDs, and list of underlying parameters are available on the CMDs/TRGB catalog of the Extragalactic Distance Database\footnote{\url{edd.ifa.hawaii.edu}} (Jacobs et al. 2009; Anand et al. 2021).

% ...........................................................................

\section{The total mass of the NGC 253 group}

A summary of data on 18 galaxies belonging to the NGC~253 group and its
vicinity is presented in Table~1. Its columns contain: (1) galaxy name; 
(2) supergalactic coordinates; (3) morphological type on the de Vaucouleurs
scale; (4) radial velocity in km\,s$^{-1}$ relative to the Local Group centroid as
given in HypeLEDA (Makarov et al. 2014, http://leda.univ-lyon1.fr) with the
standard error; (5) galaxy distance in Mpc from Extragalactic Distance
Database, EDD (Anand et al. 2021, http://edd.ifa.hawaii.edu); distances for
Scl-MM-Dw1 and Scl-MM-Dw2 were derived by Sand et al. (2014) and Toloba et al. 
(2016), respectively; (6) method used to 
determine the distance; a typical error of distance measurement via TRGB is
$\sim$5\%; we also added in the table a dwarf galaxy of low surface brightness 
``SculptorSR'' = GALEXASC J003351.79-275024.4 found by astro-amateur Sakib
Rasool (personal information); judging by its texture the galaxy is a probable
member of the NGC~253 group; (7) $K_s$- band luminosity of galaxy in the units
of solar luminosity given from Updated Nearby Galaxy Catalog (Karachentsev 
et al. 2013 (http://www.sao.ru/lv/lvgdb); (8) projected separation from NGC~253
in degrees; (9) projected searation in kpc assuming the galaxy is at the same distance as NGC~253; (10) the ``tidal index''

 $$\Theta_1 = \max [\log(L_i/D_i^3)] + C, \,\,\,\,   i = 1, 2,....N,$$
where $L_i$ is a luminosity of the neighboring galaxy in the $K$-band, $D_i$ is its 
separation from the neighbor; ranking the surrounding galaxies by
the magnitude of their tidal force, $F_i \sim L_i/D_i^3$. The most significant neighbor is the ``Main Disturber'', MD, where the constant $C$ is chosen 
so that a galaxy with $\Theta_1 = 0$ is located at the ``zero velocity sphere''
relative to its MD. Consequently, the  unrelated galaxies with a
negative $\Theta_1$ are referred to as the population of the ``general field''.
(11) name of the Main Disturber; (12) orbital mass estimate via the given
satellite, described below. All galaxies in Table~1 are ranked according
to their angular separation from NGC~253.

  As seen from Table 1, the luminosity of principal galaxy NGC~253 exceeds
the luminosity of its neighbors by more than one order of magnitude. In a 
case when a group is dominated by one massive galaxy surrounded by a set of
light test particles, one can use the ``orbital'' mass estimate (Karachentsev
\& Kudrya 2014):

 $$ M_{\rm orb} = (32/3 \pi)(1 - 2e^2/3)^{-1} G^{-1} \langle\Delta V_i^2 R_{\rm pi}\rangle.$$
Here, $G$ is the gravitation constant, e is the prevailing orbit eccentricity,
and $\Delta V$ is the radial velocity difference of a companion ``i'' at the
projected separation $R_{\rm pi}$ relative to the principal galaxy. Basing on N-body
simulations, Barber et al. (2014) estimated the typical eccentricity value
of $\langle e^2\rangle \simeq 1/2$ that yields

  $$M_{\rm orb} = (16/\pi) G^{-1} \langle \Delta V_i^2 R_{\rm pi}\rangle$$
or 
  
  $$(M_{\rm orb}/M_{\odot}) = 1.18\times 10^6 \langle\Delta V_i^2 R_{\rm pi}\rangle,$$
where $\Delta V$ and $R_{\rm pi}$ are expressed in km\,s$^{-1}$ and kpc, respectively. Individual
values of $M_{\rm orb}$ derived via different companions are given in the last
column of Table 1 in units of $10^{11} M_{\odot}$.

  The distribution of galaxies in the vicinity of NGC~253 is presented in
Fig.3 in supergalactic coordinates.  Assumed satellites of NGC~253 with
measured radial velocities are shown with solid symbols and assumed companions
without radial velocities are indicated with open circles. The field galaxies
with $\Theta_1 < 0$ are shown with crosses. For a galaxy with the luminosity
of $L_K/L_{\odot} = 10^{10.98}$, the typical virial radius is $R_v \simeq 300$~kpc, and the 
radius of zero-velocity sphere is $R_0 \simeq3.5 R_v$ (Tully 2015b). The minor and the major 
circles in Fig.3 corresponds to $R_v = 4\fdg65$  and $R_0 = 16\fdg3$. In the
sphere of the gravitational dominance of NGC~253 (i.e. inside $R_0$) there are
8 satellites with measured radial velocities. For one of them, PGC~704814,
the velocity error is too large and we ignore this galaxy. For the 7
remaining companions of NGC~253, the mean radial velocity difference is
$\langle\Delta V\rangle = -4\pm17$~km\,s$^{-1}$, the radial velocity dispersion is $\sigma_v =
42$~km\,s$^{-1}$, and the mean projected separation is $\langle R_p\rangle = 465$~kpc. The average
estimate of orbital mass via 7 satellites is $M_{\rm orb} =(8.1\pm2.6) 10^{11} M_{\odot}$,
that yields the mass-to-luminosity ratio $M_{\rm orb}/L_K = (8.5\pm2.7) M_{\odot}/L_{\odot}$.
The latter value is four times less than the ratios of ($27\pm9) M_{\odot}/L_{\odot}$
for the Milky Way and $(33\pm6) M_{\odot}/L_{\odot}$ for M~31 (Karachentsev \& Kudrya
2014).

  With such a low halo mass for NGC~253, the expected virial radius should
not be 300~kpc, but rather $\sim$200~kpc (Tully 2015a). Then only 5 satellites are inside of 
the sphere of a radius $R_0\sim 700$~kpc. For them, the dispersion of radial 
velocities, $\sigma_v = 44$~km/,s$^{-1}$, and the orbital mass estimate, $M_{\rm orb} =
(6.7\pm3.7) 10^{11} M_{\odot}$ is little changed, although the average projected 
separation of 5 satellites drops to $\langle R_p\rangle = 313$~kpc. It should be noted
that the location of these satellites appears to be very asymmetric with
respect to the principal galaxy.

  Interestingly, the field galaxies with $\Theta_1 < 0$ around NGC~253 follow
the cold unperturbed Hubble flow with the Hubble parameter $H_0 = 75$~km\,s$^{-1}$Mpc$^{-1}$ 
and $\sigma_v = 40$~km\,s$^{-1}$. According to Karachentsev et al. (2003), the Hubble
flow around NGC~253 is characterized by the zero-velocity radius $R_0 =
0.7$~Mpc, which corresponds to the total mass of the group $M_T(R_0)\simeq
6.7\times 10^{11} M_{\odot}$ in agreement with the mass estimate via internal (orbital)
motions.

  Lucero et al. (2015) performed HI observations of NGC~253 with the Karoo
Array Telescope and determined its rotation curve $V(R)$ out to the projected
separation of $\sim20$~kpc from the galaxy center. These observations show that
the rotation velocity reaches a maximum of $V_m = 214$~km\,s$^{-1}$ at $R\sim12$~kpc,
and than decreases systematically down to $\sim$185~km\,s$^{-1}$. A similar result was
obtained earlier by Hlavacek-Larrondo et al. (2011) from observations in the
H$\alpha$ and [NII] lines using Fabry-Perot interferometry. The declining
rotation curve of NGC~253 can serve as an independent indication of a small
size for the halo of this galaxy.

  NGC~253 is not the only case of a luminous spiral galaxy with a falling
rotation curve at the periphery. Casertano \& van Gorkom (1991) and
Zobnina \& Zasov (2020) identified four more such galaxies in the Local
Volume: NGC~2683, NGC~2903, NGC~3521 and NGC~5055. All of these are located
in areas of low cosmic density and have a small number of dwarf satellites.
Karachentsev et al. (2020) estimated orbital masses of these galaxies, using
radial velocities and separations of their companions. Data on 5 galaxies
in the Local Volume with declined rotation curves are presented in Table 2.
As one can see, all these spirals are characterized by high luminosities
and low relative masses of their dark halos. The average dispersion of radial velocities
of their satellites is 46~km\,s$^{-1}$ at the average projected separation of 225~kpc.
The low average value of $\langle M_{\rm orb}/L_K\rangle = (5.5\pm1.1) M_{\odot}/L_{\odot}$ is comparable
to the cosmic baryon abundance, $M_{\rm DM}/M_{\rm bar} \simeq 6$ at the stellar mass-to-$K$-
luminosity ratio of $M_*/L_K \simeq 1 M_{\odot}/L_{\odot}$ (Bell et al. 2003).

  It is implied that the galaxies with a declined rotation curve form 
a special category among the spiral galaxies of high luminosity. In the
Local Volume with the distance $D < 11$~Mpc there are 19 spiral galaxies with
the luminosity $\log(L_K/L_{\odot}) > 10.5$. If we exclude five galaxies
seen nearly face-on with indefinite rotation curves: NGC~628, IC~342, NGC~3184, 
M~101 and NGC~6946, then the relative number of cases with quasi-
Keplerian $V(R)$ is $\sim$36\%.

  Zobnina \& Zasov (2020) noted that spiral galaxies with a declined 
rotation curve do not deviate from the general Tully-Fisher relation,
$\log(L_K)$ vs. $\log(V_m)$, if $V_m$ used as an argument, and not $V(R_{\rm max})$.
Cosequently, the identification of galaxies with relatively low mass dark halos is a nontrivial
observational problem, requiring data on the kinematics of their distant
periphery.

\section{Concluding remarks}

  We measured accurate TRGB-distances of 3.62~Mpc and 3.66~Mpc for WOC2017-07 and PGC~704814, respectively, two dwarf galaxies in the vicinity
of the bright spiral galaxy NGC~253, confirming their physical association 
with NGC~253 at 3.70~Mpc. Basing on the data on radial velocities and separations of 7
salellites of NGC~253, we determined its ratio of the total (orbital) mass 
to K- band luminosity, $M_{\rm orb}/L_K = 8.5\pm2.7 M_{\odot}/L_{\odot}$, which is 3$-$4 times
less than the analogous ratio for the Milky Way or M~31. A notable feature of
NGC~253 is the presence of a descending rotation curve at the periphery.
This feature is shown by 4 more galaxies in the Local Volume: NGC~2683, NGC~2903, 
NGC~3521 and NGC~5055. All of them are characterized by a low radial
velocity dispersion of satellites, $\sigma_v = (42 - 54)$~km\,s$^{-1}$, and a low mass
of dark halo, $\log(M_{\rm orb}/M_{\odot}) = 11.09 - 11.91$. The mean total mass-to-$K$-
luminosity ratio for them is ($5.5\pm1.1) M_{\odot}/L_{\odot}$ on the scale of $\sim225$~kpc.

  It is suggested that there is a special population of galaxies with
quasi-Keplerian rotation curves, which are found mainly in regions of low
cosmic density. In the Local Volume, their relative number is $\sim$1/3 among
luminous spiral galaxies. The dark halos of these galaxies appear to be restricted in mass an extent.

  Recently, Correa \& Schaye (2020) and Seo et al. (2020) studied the
dark-to-stellar mass ratio, $M_{\rm DM}/M_*$, depending on the morphology of
galaxies from Sloan Digital Sky Survey. Both teams reported systematically
lower values of $M_{\rm DM}/M_*$ for disc-dominated (blue) galaxies than bulge-
dominated (red) ones. A similar effect was found for 2MASS isolated galaxies
by Karachentseva et al. (2011). Correa \& Schaye (2020) explain this difference
by the assumption that the stellar discs are more massive because they had 
more time for gas accretion and star formation. Also, according to Seo et al.
(2020), ``the system velocity dispersion of satellite galaxies show a remarkably
tight correlation with the central velocity dispersion of their primary
galaxies for both red and blue samples.'' The study of the relationship between
the kinematics of satellites and inner kinematics of isolated luminous galaxies
seems to be an important observational problem.

   {\bf Acknowledgements.}  This work is based on observations made with the NASA/ESA
Hubble Space Telescope. STScI is operated by the Association of Universities 
for Research in Astronomy, Inc. under NASA contract NAS 5-26555. IDK is
suported by RNF grant 19-12-00145.

                    {\bf References}

Anand, G.S., Rizzi, L., Tully, R.B., et al., 2021, in preparation

Arp H., 1985, Astron. J. 90, 1012

Barber C., Starkenburg E., Navarro J.F., et al, 2014, MNRAS, 437, 959 

% Battaglia G., Fraternali F., Oosterloo T., Sancisi R., 2006, A\&A, 447, 49

Bell E.F., McIntosh D.H., Katz N., Weinberg M.D., 2003, ApJ Suppl., 149, 289

Cannon J.M., Dohm-Palmer R.C., Skillman E.D., et al, 2003, AJ, 126, 2806

Casertano S., van Gorkom J.H., 1991, AJ, 101, 1231

Colless, M., Peterson, B.A., Jackson, C.A. et al. 2003, arXiv:2003.6581

Correa C.A., Schaye J.,  2020, MNRAS, 499, 3578

de Vaucouleurs G., 1959, Astrophys. J. 130, 718 

Dolphin, A.E. 2000, PASP, 112, 1383

Dolphin, A.E. 2016, DOLPHOT: Stellar photometry, ascl:1608.013

Hlavacek-Larrondo J., Carignan C., Daigle O., et al, 2011, MNRAS, 411, 71

Jacobs, B.A., Rizzi, L., Tully, R.B., et al., 2009, AJ, 138, 332

Jarrett, T. H.; Chester, T.; Cutri, R.; Schneider, S. E.; Huchra, J. P., 2003, Astron. J. 125, 525

Jerjen H., Freeman K.C., Binggeli B., 1998, Astron. J. 116

% Karachentsev I.D., Makarova L.N., Tully R.B., et al, 2020, A\&A, 643A, 124

Karachentsev I.D., Neyer F., Spani R., Zilch T., 2020, AN, 341, 1037

Karachentsev I.D., Kudrya Y.N., 2014, AJ, 148, 50

Karachentsev I.D., Makarov D.I., Kaisina E.I.2013, Astron. J., 145, 101

Karachentsev I.D., Grebel E.K. Sharina M.E., et al, 2003, A\&A, 404, 93

Karachentseva V.E., Karachentsev I.D., Melnyk O.V., 2011, AstBu, 66, 389

Koribalski, B.S., Staveley-Smith, L., Kilborn, V.A. et al., 2004, Astron. J., 128, 16

Lucero D.M., Carignan C., Elson E.C., et al, 2015, MNRAS, 450, 3935 

Makarov, D.I., Makarova, L.N., Rizzi, L. et al., 2006, AJ, 132, 2729 

Makarov, D.I., Prugniel P., Terekhova N., et al, 2014, A\&A, 570A, 13

McQuinn, K.B.W., Skillman, E.D., Dolphin, A.E., Berg, D., Kennicutt, R., 2017, AJ, 154, 51

Rizzi, L., Tully, R.B., Makarov, D.I. et al. 2007, ApJ, 661, 815

Sand D.J., Crnojevich D., Strader J., et al, 2014, ApJ, 793L, 7

Schlafly, E.F., Finkbeiner, D.P. 2011, ApJ, 737, 103

Seo G., Sohn J., Lee M.G., 2020, ApJ, 903, 130

Toloba E., Sand D.J., Spekkens K., et al, 2016, ApJ, 816L, 5 

 Tully R.B., 1988, Nearby Galaxy Catalog, Cambridge Univ. Press 

Tully, R.B., 2015a, Astron. J. 149, 54

Tully, R.B., 2015b, Astron. J. 149, 171

Westmeier T., Obreschkow D., Calabretta M., et al, 2017, MNRAS, 472, 4832

Wu, P.-F., Tully, R.B., Rizzi, L. et al. 2014, AJ, 148, 7

Zobnina D.I., Zasov A.V., 2020, Astronomy Reports, 64, 295

\newpage

  \clearpage  
\begin{figure}
\plotone{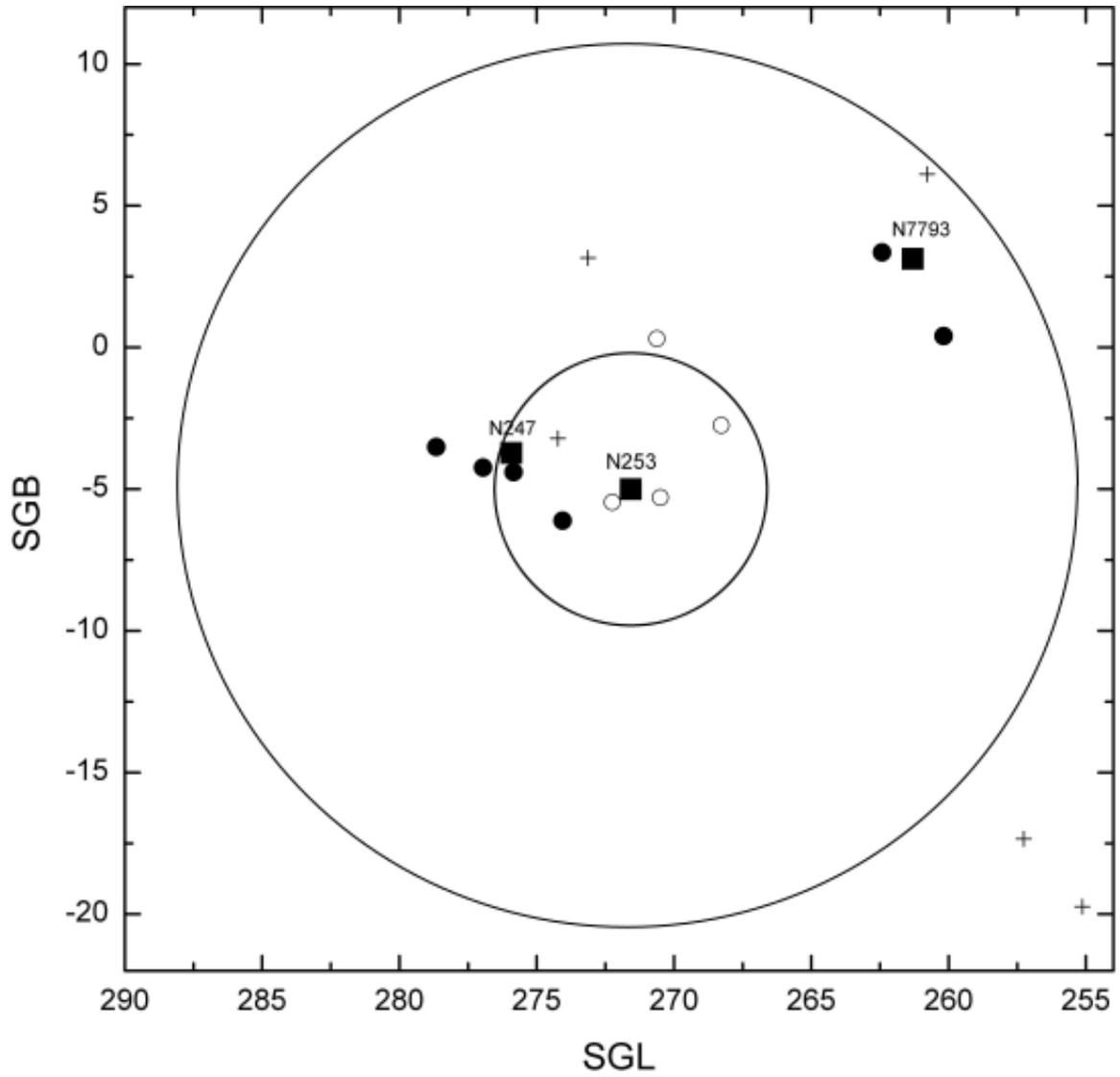}
\caption{Distribution of assumed NGC~253 satellites (squares, circles) and background
       galaxies (crosses) in supergalactic coordinates. Assumed satellites
       without radial velocities are indicated with open circles. The virial
       radius of 300~kpc ($4\fdg65$) and the zero-velocity radius of 1~Mpc ($16\fdg$)
       are shown with minor and major circles.}
\end{figure}

%\end{document}
 \begin{table}
\caption{Galaxies in and around the NGC 253 group.}
\begin{tabular}{lcrcclrrrrlc} \hline

  Name       &  SGL   SGB  &   T &  $V_{\rm LG}\pm$ e  & D   & meth & $\log L_K$ & $r_p$ &  $R_p$ & $\Theta_1$ & MD  & $M_{\rm orb}$\\
\hline
   (1)      &      (2)      & (3) &  (4)    &  (5) &  (6) &  (7)   &   (8)  & (9)  & (10) & (11)   &(12)\\
\hline
 NGC 253    &  271.57$-$05.01 &  5    & 276$\pm$2  & 3.70 & trgb & 10.98  &    0   &   0  &  0.1 & N 247  & $-$ \\
 Scl-MM-Dw2 &  272.25$-$05.47 & $-$2  &   $-$      & 3.12 & trgb &  7.36  &   0.82 &  53  &  0.7 & N 253  & $-$ \\
 Scl-MM-Dw1 &  270.50$-$05.30 & $-$2  &   $-$      & 3.94 & trgb &  6.77  &   1.11 &  72  &  1.8 & N 253  & $-$  \\
 WOC2017-07 &  274.06$-$06.12 & 10    & 288$\pm$ 5 & 3.62 & trgb &  6.17  &   2.73 & 176  &  2.1 & N 253  & 0.3\\
 DDO 226    &  274.23$-$03.21 & 10    & 412$\pm$ 2 & 4.92 & trgb &  7.71  &   3.21 & 207  &$-$0.3& N 253  & $-$ \\
 SculptorSR &  268.28$-$02.75 & 10    &   $-$      & 3.70 & mem  &  6.36  &   3.99 & 258  &  1.2 & N 253  & $-$  \\
 DDO 6      &  275.84$-$04.40 & 10    & 347$\pm$ 2 & 3.43 & trgb &  7.08  &   4.31 & 278  &  1.3 & N 253  &16.5 \\
 NGC 247    &  275.92$-$03.73 &  7    & 210$\pm$ 2 & 3.71 & trgb &  9.50  &   4.53 & 293  &  1.6 & N 253  &15.1\\
 Sc 22      &  270.62$+$00.31 & $-$3  &   $-$      & 4.29 & trgb &  7.15  &   5.40 & 349  &  0.5 & N 253  & $-$   \\
 ESO 540-032&  276.95$-$04.24 & 10    & 285$\pm$ 7 & 3.63 & trgb &  6.83  &   5.43 & 351  &  1.4 & N 247  & 0.3\\
 KDG 2      &  278.66$-$03.52 & 10    & 290$\pm$ 7 & 3.56 & trgb &  6.85  &   7.24 & 468  &  1.0 & N 253  & 1.1\\
 NGC 59     &  273.13$+$03.16 & $-$3  & 431$\pm$ 2 & 4.90 & trgb &  8.66  &   8.32 & 537  &$-$0.4& N 253  & $-$   \\
 PGC 704814 &  262.42$+$03.35 & 10    & 299$\pm$89 & 3.66 & trgb &  6.90  &  12.39 & 800  &  2.1 & N7793  &(4.2)\\
 ESO 349-031&  260.18$+$00.40 & 10    & 234$\pm$ 3 & 3.21 & trgb &  7.12  &  12.61 & 814  &  0.2 & N 253  &17.0 \\
 NGC 7793   &  261.30$+$03.12 &  6    & 250$\pm$ 2 & 3.71 & trgb &  9.70  &  13.10 & 846  &  0.2 & N 253  & 6.7 \\
 UGCA 442   &  260.78$+$06.11 &  8    & 300$\pm$ 6 & 4.36 & trgb &  8.03  &  15.49 &1000  &$-$0.3& N 253  &  $-$ \\
 NGC 625    &  257.27$-$17.34 &  8    & 320$\pm$ 6 & 4.02 & trgb &  8.96  &  18.88 &1219  &$-$0.3& N 253  &  $-$ \\
 ESO 245-005&  255.13$-$19.74 &  9    & 307$\pm$ 2 & 4.57 & trgb &  8.53  &  24.39 &1575  &$-$0.7& N 253  &  $-$ \\
\hline
\end{tabular}
{{\bf Note}: (1) galaxy name; 
(2) supergalactic coordinates; (3) morphological type on the de Vaucouleurs
scale; (4) radial velocity in km\,s$^{-1}$ relative to the Local Group centroid; (5) galaxy distance in Mpc; (6) method used to 
determine the distance;  (7) $K_s$- band luminosity of galaxy in the units
of solar luminosity; (8) projected separation from NGC~253
in degrees; (9) projected separation in kpc; (10) the ``tidal index'' $\Theta_1$;
(11) name of the Main Disturber; (12) orbital mass estimate via the given
satellite.}
\end{table}
 %\end{document}
 \begin{table}
 \caption{Luminous galaxies in the Local Volume with declined rotation curves.}
  \begin{tabular}{lcrcccrcc} \hline

 Name    & Type &   D &  $\log L_K $& $n_v$ & $\sigma_v$& $\langle R_p\rangle$& $\log M_{\rm orb}$  &$M_{\rm orb}/L_K$\\
\hline
 NGC 253   & 5  &  3.70 & 10.98  &  7 &    42 &  465   &  11.91  &   8.5$\pm$2.7  \\
 NGC2683   & 3  &  9.82 & 10.81  &  2 &    43 &   49   &  11.09  &   1.9$\pm$1.3  \\
 NGC2903   & 4  &  8.87 & 10.82  &  4 &    45 &  198   &  11.68  &   7.3$\pm$6.4  \\
 NGC3521   & 4  & 10.70 & 11.09  &  2 &    46 &  198   &  11.77  &   4.8$\pm$4.0  \\
 NGC5055   & 4  &  9.04 & 11.00  &  4 &    54 &  216   &  11.71  &   5.1$\pm$1.8  \\
  \hline
 Mean      & 4  &  8.43 & 10.94  &  4 &    46 &  225   &  11.63  &   5.5$\pm$1.1  \\
\hline
\end{tabular}

{{\bf Note}: (1) galaxy name; 
(2) morphological type on the de Vaucouleurs
scale; (3) galaxy distance in Mpc; (4) $K_s$- band luminosity of galaxy in the units
of solar luminosity; (5) number of companions with measured velocities; (6) velocity dispersion of companions in km s$^{-1}$; (7) mean separation of companions in kpc; (8) log mass of group from orbital dynamics in solar masses; (9) ratio of orbital mass to K band luminosity in solar units
}
\end{table}
    \end{document}